\documentclass[12pt,preprint]{aastex}

\def\ni{\noindent}
\def\txg{t_{x\gamma,4}}
\def\lta{\lower2pt\hbox{$\buildrel {\scriptstyle <} 
   \over {\scriptstyle\sim}$}}
\def\gta{\lower2pt\hbox{$\buildrel {\scriptstyle >} 
   \over {\scriptstyle\sim}$}}

\newcount\eqnumber
\def\chaphead{}

\def\new{\hbox{\chaphead\the\eqnumber}\global\advance\eqnumber by 1}
\def\ref#1{\advance\eqnumber by -#1 \chaphead\the\eqnumber
     \advance\eqnumber by #1 }
\def\first{\hbox{(\chaphead\the\eqnumber{a}}\global\advance\eqnumber by 1}
\def\last{\advance\eqnumber by -1 \hbox{\chaphead\the\eqnumber}\advance
     \eqnumber by 1}
\def\eq#1{\advance\eqnumber by -#1 equation (\chaphead\the\eqnumber
     \advance\eqnumber by #1}
\def\eqnam#1{\xdef#1{\chaphead\the\eqnumber}}

\begin{document}

\title{ X-ray Lines From Gamma-ray Bursts}
\author{Pawan Kumar}
\affil{Astronomy Department, University of Texas, Austin, TX 78731}
\author{Ramesh Narayan}
\affil{Dept. of Astrophysical Sciences, Princeton University, Princeton, 
       NJ 08544}
\affil{and Harvard-Smithsonian Center 
    for Astrophysics, 60 Garden Street, Cambridge, MA 02138}
\email{pk@astro.as.utexas.edu, rnarayan@cfa.harvard.edu}

\begin{abstract}

X-ray lines have been recently detected in the afterglows of a few
gamma-ray bursts.  We derive constraints on the physical conditions in
the line-emitting gas, using as an example the multiple K$_\alpha$
lines detected by Reeves et al. (2002) in GRB 011211.  We argue that
models previously discussed in the literature require either a very
extreme geometry or too much mass in the line-emitting region.  We
propose a new model in which gamma-rays and radiation from the early
x-ray afterglow are back-scattered by an electron-positron pair screen
at a distance of about $10^{14}-10^{15}$ cm from the source and
irradiate the expanding outer layers of the supernova ejecta, thereby
producing x-ray lines.  The model suffers from fewer problems compared
to previous models.  It also has the advantage of requiring only a
single explosion to produce both the GRB and the supernova ejecta, in
contrast to most other models for the lines which require the
supernova to go off days or weeks prior to the GRB.  The model,
however, has difficulty explaining the $>10^{48}$ ergs of energy
emitted in the x-ray lines, which requires somewhat extreme choices of
model parameters.  The difficulties associated with the various models
are not particular to GRB 011211.  They are likely to pose a problem
for any GRB with x-ray lines.

\end{abstract}

 
\section{Introduction}

Although much has been learned about gamma-ray busts (GRBs) in the
last several years, the nature of the central exploding object remains
uncertain (see Piran, 2000; van Paradijs et al. 2000; Meszaros, 2001
for recent reviews).  The possible detection of x-ray lines in GRB
afterglows is very important in this connection since the lines
provide strong constraints on models.

Piro et al. (2000) claimed to detect an Fe K$_\alpha$ line and the
corresponding recombination edge in GRB 991216.  The line was seen
about 1.5 days (in the observer frame) after the initial burst.  There
have been claims for iron lines also in a few other bursts (GRB
970508: Piro et al. 1999; GRB 970828: Yoshida et al. 1999; GRB 000214:
Antonelli et al. 2000).

Recently, Reeves et al. (2002) reported the detection of x-ray
K$_\alpha$ lines of Mg XI (or XII), Si XIV, S XVI, Ar XVIII and Ca XX
in GRB 011211.  The presence of multiple lines at apparently the same
redshift makes this detection particularly interesting.  If the lines
are emitted isotropically, GRB 011211 must have emitted over $10^{48}$
ergs in each line as measured in the source frame.  The emission
occurred a few hours after the GRB and continued for about half an
hour in the source frame.  Since the lines are blue-shifted by
$v/c\sim0.1$ with respect to the GRB, the emission must have occurred
within a sub-relativistic outflow from the source.

We show in this paper that it is quite challenging to come up with a
viable theoretical model to explain the lines seen in GRB 011211.
Recently, Borozdin \& Trudolyubov (2002) have suggested that the lines
may be an instrumental artifact, which would of course solve the
theoretical problems.  In the rest of the paper we derive constraints
on some of the popular models that have been proposed so far for the
line emission.  The constraints apply to GRB 011211 and also to any
other GRB for which a line with an energy output of about 10$^{48}$
ergs is detected; this level of line emission roughly corresponds to
the detection threshold of the current generation of instruments for a
high redshift GRB.  We also propose in this paper a new model for line
production, which we believe suffers from fewer problems than the
previous models.

Two broad classes of models have been proposed to explain the x-ray
lines.  In one model, it is assumed that the GRB is associated with a
long-lived engine, with a duration of several hours to a day (Meszaros
\& Rees 2000).  Part of the long-duration emission from the engine is
intercepted by a funnel that has been carved out of the supernova
ejecta.  The intercepted radiation is reprocessed into x-ray line
photons by the photoionized gas in the surface layers of the funnel.
Lazzati, Ramirez-Ruiz \& Rees (2002) have shown that this model is
ruled out in the case of GRB 011211, the reason being that the
radiation from the central engine which causes the line emission
should also be visible to the observer.  There should be a very
significant level of continuum radiation, which was not observed.

In the second model, it is proposed that the GRB explosion is
preceded, by roughly a week or so, by a supernova explosion which
ejects a dense slab of gas moving at a speed $v\sim 0.1c$.  The
ejected material reaches a distance $> 10^{15}$ cm by the time the GRB
goes off.  The x-ray photons from the GRB and early afterglow are
intercepted by the slab and the radiation is reprocessed into line
photons (Vietri \& Stella, 1998; Vietri et al. 2001).  The geometry of
the outflowing gas is somewhat flexible.  It could be a
quasi-spherical shell, as in the model suggested by Reeves et
al. (2002), or it could be in the form of a large funnel carved out in
the previously ejected material.  In either case, the delay of a few
hours between the GRB and the line emission is explained as the
geometrical time delay associated with the extra path length that
photons have to travel from the GRB to the photoionized layer and
then to the observer.

A third possibility, which we elaborate upon here, is that the
supernova and the GRB go off simultaneously, but that the radiation
from the GRB and early afterglow is scattered by a nearby scattering
medium and then intercepted by the expanding surface of the supernova
ejecta.  The lines are produced from the surface layers of the ejecta.
The time delay in this case measures the path length from the GRB to
the scatterer and back to the ejecta.  The basic outline of this model
was briefly described by Vietri et al. (2001), who however considered
the model implausible.  We conclude just the opposite in this paper.
We argue that, if the lines in GRB 011211 are real, they were probably
produced by something resembling this model.  Vietri et al. (2001) did
not specify the nature of the scatterer.  In our model, the scattering
occurs in a pair screen created by the gamma-ray photons from the GRB.
An attractive feature of the model is that the radius at which the
pair screen is likely to form is close to what is needed to explain
the observed time delay.

We provide some general constraints on the various models in \S2 and
find serious problems with all proposals.  We show in \S3 that the
pair screen scattering model can account for the x-ray lines in GRB
011211 provided there is sufficient ambient density in the vicinity of
the GRB to scatter gamma-rays to produce a significant pair screen,
and provided the surface layers of the exploding star move with a
significant speed $\sim0.3c$. We conclude with a summary and
discussion in \S4.  Some of the radiation physics relevant to line
emission is discussed in the Appendix A.
 
\section{General Constraints}

Consider a photoionized medium with a radius $R=10^{15}R_{15}$ cm as
seen projected on the plane of the sky (i.e., perpendicular to the
line of sight from the observer).  Let the line-emitting gas have a
thickness $d$ and optical depth $\tau$ parallel to the line of sight.
Let the electron density be $n_e=10^{15}n_{15} ~{\rm cm^{-3}}$, the
temperature be $10^7T_7$ K, and let the number of electrons per Si ion
be $4\times10^4a_s$; we focus on Si because it is the strongest line
seen in GRB 011211.  The parameter $a_s$ is defined such that $a_s=1$
corresponds to solar composition and $a_s<1$ corresponds to a medium
in which the heavy elements are enhanced by a factor $1/a_s$.

Let the medium be irradiated by x-ray continuum radiation from the GRB
for a time $t=10^2t_2$ s.  The efficiency for converting x-ray
continuum radiation (in the 1--10 keV band) to x-ray line radiation is
typically a few percent under optimal conditions (Lazzati et
al. 2002).  Therefore, the energy in continuum irradiation is $\sim 50
\zeta_{50} E_x$, where $E_x=10^{48}E_{48}$ ergs is the energy emitted
in line photons.  The continuum photon number flux incident on the
line producing region is thus
$$ 
{\aleph}_x = {50 \zeta_{50} E_x\over 4\pi R^2 t \epsilon_\alpha}
\sim 10^{25} {\zeta_{50}E_{48}\over R^2_{15} t_{2}\epsilon_3} \; {\rm cm}^{-2}
{\rm s}^{-1},
$$ 
where $\epsilon_x=3\epsilon_3$ keV is the energy of each line photon.

The photoionization cross-section for hydrogenic ions is
$\sigma_\gamma = 7.9\times10^{-18} z^{-2}$ cm$^{2}$, where $z$ is the
atomic number of the line-emitting species ($z=14$ for Si).  For the
photon number flux estimated above, we see that the photoionization
time for Si XIII is about 3x10$^{-6}$ s.  In comparison, the
recombination time for Si XIV is
$$
t_{rec} \sim 5.9\times10^{-4}{T_{7}^{1/2} \over z_{14}^2 n_{15}} ~{\rm
s},
$$
which, for an electron density of 10$^{15}$ cm$^{-3}$, is about $10^3$
times longer than the photoionization time.  In the following, we
therefore assume that the line emission is controlled by the
recombination rate.

The rate at which Si K$_\alpha$ photons are emitted by
the gas per unit projected area is
\begin{eqnarray}
\dot N_{Si}  = {n_{Si}d\over t_{rec}}=\left({n_{e}\over4\times10^4a_s}
\right) \left({\tau\over\sigma_T n_e}\right) 
\left({z_{14}^2n_{15}\over5.9\times10^{-4}T_7^{1/2}}\right) \nonumber \\
  = 6.4\times10^7{z_{14}^2\tau n_e\over a_s T_7^{1/2}} ~{\rm
cm^{-2}s^{-1}}.\quad\quad\quad\quad\quad\quad\quad \nonumber
\end{eqnarray}

The line flux is
$$
F_{Si} = \epsilon_\alpha\dot N_{Si} =
3.1\times10^{14}{z_{14}^2\tau\epsilon_3 n_{15}\over a_s T_7^{1/2}}
~{\rm ergs\,cm^{-2}s^{-1}}.  \eqno(\new)
$$ 
The irradiating continuum flux in 1--10 keV photons is
$F_{1-10}=50\zeta_{50} F_{Si}$, and so the ionization parameter is
$$
\xi = {4\pi F_{1-10}\over n_e} = 190 {z_{14}^2\epsilon_3\zeta_{50}
\over T_7^{1/2}} {\tau\over a_s}. \eqno(\new) 
$$
For $T_7\sim1$, $\tau\sim1$, $a_s\sim 0.1$ (ten times solar abundance,
see Reeves et al. 2002 and Appendix A3), we find $\xi\sim2\times
10^3$, which is approximately the value needed for strong line
emission according to Lazzati et al. (2002) for a ten times solar
composition gas.  Our analysis in Appendix A2 also gives a similar
requirement for $\xi$.  Thus, it appears that an optimum ionization
parameter is obtained quite naturally in this problem using reasonable
parameters.

Since each gas element in the medium is irradiated and emits lines for
a time $10^2t_2$ s, the total energy emitted in Si lines is
$$
E_{Si} = F_{Si}\pi R^2 10^2t_2 = 5\times10^{47} {\xi_3 t_2n_{15}
R_{15}^2\over\zeta_{50}} ~{\rm ergs},
$$
where we have written the answer in terms of $\xi_3=\xi/10^3$.  For an
observed line energy of $10^{48}E_{48}$ ergs, we then find
$$
n_{15} \sim {2 \zeta_{50} E_{48}\over \xi_3 t_2R_{15}^2}. \eqno(\new) 
$$
Note that a number of parameters have disappeared from the final
expression for $n_e$.  This is because we have expressed the result in
terms of the ionization parameter.

Equation (3) is an important constraint on the density of the
radiating layer.  The constraint applies not only to the Si line
explicitly considered above, but also to lines of Fe and other
elements of similar abundance when appropriate values for $E_{48}$ and
$\xi_3$ are substituted.  As an example, for GRB 991216, Piro et
al. (2000) reported the detection of an Fe K$_\alpha$ line with
$E_{48}\sim 10$. Since the optimum $\xi_3$ for producing the Fe line
is typically about 10 times larger than for the Si line (see Fig. 1
and also Lazzati et al. 2002), the numerical coefficient in equation
(3) remains unchanged for the particular claimed Fe line.

The constraint expressed by equation (\last), coupled with another 
constraint which comes from the observed time delay (see below), severely 
limits possible models for the line emission.
We now consider three very different models for the line emission.

\medskip
\noindent{\bf 1. Shell Model}:
We assume that the radiating gas is in the form of a quasi-spherical
shell at a radial distance $D=10^{15}D_{15}$ cm from the GRB engine.
Let the radiation from the GRB be beamed within a cone of half-angle
$0.1 \theta_{-1}$ radians.  The observed time delay of $10^4\txg$ s
between the GRB and the X-ray lines requires
$$ 
R_{15} ={6\txg\over\theta_{-1}}, \qquad D_{15} =
{60\txg\over\theta_{-1}^2}. \eqno(\new)
$$
Given this estimate of $D$ and assuming a beaming angle of 0.1
radians, we note that a shell moving at speed $0.1c$ must have been
ejected nearly a year before the GRB.  The time difference may be
reduced to a few days if the beam is significantly wider (Reeves et
al. 2002).  Substituting the above estimate of $R_{15}$ into equation
(3) we find that the mean electron density in the shell is
$$
n_{15} \sim 0.06{\theta_{-1}^2\zeta_{50}E_{48}\over\xi_3 t_2\txg^2},
\eqno(\new)
$$
and the thickness of the shell is
$$
d = {\tau\over\sigma_Tn_e}=2.7\times10^{10}{\tau\xi_3 t_2\txg^2
\over\zeta_{50}\theta_{-1}^2E_{48}}\;{\rm cm},
$$
$$
{d\over D} = 4.5\times10^{-7}{\tau\xi_3 t_2\txg
\over\zeta_{50}E_{48}}.
$$
We consider such a small value of $d/D$ to be highly unlikely.
Because in this model the irradiation occurs on the inside of the
shell and the observed line emission is from the outside, the quantity
$d$ directly measures the total thickness of the shell.  Even if the
shell was initially ejected with zero thickness, we estimate that
thermal expansion would have broadened it considerably.  For instance,
if the shell was created with a temperature of 10$^6$ K (a
conservative estimate) and it subsequently cooled adiabatically, then
the relative expansion velocities of the two surfaces of the shell
would be about 100 km s$^{-1}$, which is to be compared with the mean
velocity of $0.1c$.  It is unlikely that $d/D$ would be less than the
ratio of the two velocities, which gives a limit $d/D\gta3\times
10^{-3}$, orders of magnitude larger than the value obtained in the
shell model.  Note that there is very little leeway in the values of
the parameters $\tau$, $\xi_3$, $t_2$, $\zeta_{50}$ and $E_{48}$ in
the expression for $d/D$.  As discussed above, a similar constraint
should apply also to the Fe line observation of Piro et al. (2000),
except that the numerical coefficient is a little different because of
the longer time delay in that case.  We conclude that the shell model
is not viable.

\medskip
\noindent{\bf 2. Funnel Model}: We assume that the radiating layer is
on the inside of a conical or quasi-cylindrical funnel with an opening
angle $0.1\theta_{-1}$ radians (as viewed from the GRB engine).  The
estimates of $R_{15}$, $D_{15}$ and $n_{15}$ obtained in equations
(4), (5) for the shell model are valid in this case as well.  (Recall
that $R_{15}$ refers to the projected radius of the emitting region,
and that $\tau$ refers to the optical depth parallel to the line of
sight, not perpendicular to the inclined funnel surface.)  In this
model there is no longer any difficult with having a very small value
of $d/D$.  The ionizing photons from the GRB impinge on the inner
surface of the funnel and the line photons are also emitted from the
same surface.  Thus, the radiating region can be an arbitrarily thin
layer (unlike in the case of the shell model).

There is, however, a problem with the amount of mass required in the
model.  Let us assume that the funnel is carved out of a roughly
quasi-spherical external medium (most probably the supernova ejecta).
Given the electron density $n_e$ and the radius of the sphere $D$, we
estimate the external mass to be
$$
M \sim {4\pi\over3}D^3n_em_p \sim 4.3\times10^7
{\zeta_{50}\txg\over\xi_3 t_2\theta_{-1}^4}M_\odot.
$$
For $\theta_{-1}\sim1$, the mass is unacceptable large.  Even if we
take $\theta_{-1}\sim10$, i.e., no beaming, the mass is still much too
large.  As in the case of the shell model, we do not have much freedom
in choosing the other parameters.

One way to decrease the mass requirement in the funnel model is to
enhance the density in the funnel wall relative to the rest of the
ejecta.  For example, one could imagine that initially there was no
funnel, and that it was the GRB itself that pushed the material aside
to form the funnel.  It is conceivable that the material pushed aside
could have piled up on the walls, giving an enhanced density there.
It is not clear that this mechanism can produce the orders of
magnitude density enhancement needed to reduce the mass estimate to a
reasonable value.

\medskip
\noindent{\bf 3. Scattering Screen Model}: 
In this model, the photoionization occurs indirectly.  Photons from
the GRB travel out to a screen, are scattered, and then irradiate the
line-emitting gas, which is located in the surface layers of the
supernova ejecta.  Because the geometry is very different from the
previous two models, there are two important changes.  First, the time
delay of $10^4$ s between the GRB and the line emission measures the
distance to the scattering screen ($> 10^{14}$ cm) but not the size of
the line-emitting ejecta.  Therefore, the estimates given in equation
(4) are not valid.  Instead, if we assume that the ejecta have
expanded at speed $0.1c\beta_{*,-1}$ for $10^4\txg$ s, we estimate that $R\sim
D\sim 10^{13.5}\txg\beta_{*,-1}$ cm.  Second, the irradiation occurs for a time
$\sim10^4\txg$ s rather than $10^2$ s, and so we expect $t_2\sim100\txg$.
Both changes help to ease some of the constraints.

With the above choices for $R$ and $t$, we find
$$ n_{15} = 20 {\zeta_{50}E_{48}\over\xi_3 \txg^3 \beta^2_{*,-1}}. 
    \eqno(\new) $$
The mass in the ejecta is then
$$
M \sim {4\pi\over3}R^3n_em_p \sim 2{\zeta_{50}E_{48}\beta_{*,-1}
\over \xi_3} M_\odot,
$$
and the kinetic energy is
$$
E_{KE} \sim M(0.1c\beta_{*,-1})^2/2 = 2\times10^{52}{\zeta_{50}E_{48}
   \beta^3_{*,-1} \over \xi_3} ~{\rm ergs}.
$$
The kinetic energy may be overestimated by a factor of a few because
the velocity of $0.1c$ presumably refers only to the surface layers,
while the bulk of the ejecta may be moving substantially slower.
Overall, the parameters are more reasonable in this model than in the
previous two models.  The mass and energy requirements, for instance,
are consistent with an underlying supernova.

Two issues remain to be clarified in this model, both of which are
discussed in \S3.  First, we need to identify the nature of the
scattering screen.  We suggest that it is a pair plasma cloud which is
temporarily created by the GRB radiation itself.  Second, we have to
check whether there is enough scattered radiation to power the rather
impressive level of line emission seen in GRB 011211.

\section{X-ray Reflection from a Pair Screen}

As the pulse of gamma-ray photons from a GRB moves outward, some of
the photons will be scattered by electrons in their path and will
collide with other outgoing photons to produce electron-positron
pairs.  This process has been considered by a number of authors in the
context of GRBs (Thompson \& Madau 2000; Meszaros et al. 2000;
Beloborodov 2002; Kumar \& Panaitescu 2002).

An exponential growth of pairs occurs when the optical depth for a
scattered GRB photon to pair-produce off outgoing Mev photons is
greater than 1.  This condition gives an upper limit on the distance
of the pair screen from the central explosion: $R_\pm\sim 10^{15}
E_{52}^{1/2}$ cm.  Here, $E_{52}$ measures the isotropic equivalent
energy in the gamma-ray pulse in photons of energy greater than 1 Mev,
where $E_{52}\approx E_{GRB} (2m_e c^2/h\nu_p)^{-\beta}/10^{52} ~{\rm
ergs}$, $E_{GRB}$ is the total energy in the GRB pulse, $\nu_p$ is the
frequency at the peak of the spectrum, and $\beta$ is the high energy
spectral index.  Approximately 10$^3$ pairs can be created per ISM
electron when conditions are right, but the number can be much less in
other situations.  If the optical depth of the pair screen becomes
$\sim 1$, further pair creation is expected to be quenched and so the
optical depth will not increase much above unity.

Let the medium in the vicinity of the GRB explosion be stratified with
density decreasing as $r^{-2}$, as appropriate for a wind from the
pre-supernova star: $n(r)= n_0 (r/r_0)^{-2}$.  For a mass loss rate of
10$^{-5}$ yr$^{-1}$, which is average for a Wolf-Rayet star, and a
wind speed of 10$^3$ km s$^{-1}$, the number density at a radius of
10$^{15}$ cm is about 10$^6$ cm$^{-3}$.  The optical depth in the wind
is dominated by the radius at which the GRB photons are produced
(where the density is highest).  In the internal shock model for GRBs
this radius is a $r_1\sim {\rm few}\times 10^{14}$ cm.  Let us assume
that the pair screen forms between $r_1$ and an outer radius $r_2$.
Let us also imagine that, on average, each ISM electron produces
$\eta_\pm$ pairs.  The optical depth of the pair screen is then
$$ 
\tau = n_0 \sigma_T \eta_\pm r_0^2/r_1, 
$$ 
Assuming a mass $2m_p$ per ISM electron, the mass of the screen is
$$ 
M = \pi\theta_j^2 2 m_p n_0 r_0^2 r_2, 
$$
where $\theta_j$ is the jet opening angle.  

Let the pair screen move with Lorentz factor $\gamma_\pm$ as a result
of absorbing momentum from the GRB photons.  The momentum of the
screen may be written in the form
$$ 
P = {2\pi\theta_j^2 \tau r_2 r_1 m_p c\gamma_\pm\over \sigma_T \eta_\pm}.
$$
The momentum taken out of the GRB pulse is
$E_{GRB}\times\min(\tau,1)/(c\gamma_\pm^2)$, where the factor
$\gamma_\pm^{-2}$ allows for the fact that, for a stationary observer,
the scattered photons move at an angle of $1/\gamma_\pm$ with respect
to the radial direction.  Equating the momentum taken out of the GRB
pulse to $P$ we obtain an estimate for the Lorentz factor of the pair
screen:
$$
  \gamma_\pm = \left[ {\sigma_T E_{GRB}\eta_\pm \min(\tau,1)\over
      2\pi\theta_j^2 m_p r_2 r_1 c^2 \tau}\right]^{1/3} =
       \left( { 0.07 E_{51}\eta_\pm\min(\tau,1)\over\theta_j^2 r_{1,15} 
       r_{2,15}\tau}\right)^{1/3}.
$$

For $E_{51}=0.5$, $\theta_j=0.2$, $\eta_\pm=10$,
$r_{1,15}=r_{2,15}/4=0.5$ and $\tau\le1$ we find $\gamma_\pm=2.1$.
The total energy scattered backward is about a factor of 18 less than
the energy in the original GRB pulse because of Doppler
de-amplification.  However, the peak of the spectrum of the
back-scattered radiation is a factor of $\sim4$ smaller than the peak
of the GRB spectrum.  If the latter is at 100 kev, we expect the
ejecta from the exploding star to be bathed in a radiation field with
a peak at $\lta 25$ kev, which is closer to the optimum energy for
photoionizing the gas in the ejecta.

In order to have significant back-scattering of photons, $\tau$ should
be close to unity.  Taking $\eta_\pm=10$, $r_0=10^{15}$ cm,
$r_1=5\times10^{14}$ cm, this requires $n_0 \sim 7\times10^7 ~{\rm
cm^{-3}}$, which is substantially larger than the number density
expected for a typical WR star wind.  The pre-supernova star must have
had an unusually heavy mass-loss rate within the last few years of its
life just before the explosion.

Assuming $\tau\ \gta\ 1$, the fraction of the backward scattered
energy intercepted by the exploding star, assuming the ejecta are
expanding uniformly with a velocity 0.15c,\footnote{$^1$}{For a
spherically expanding shell with speed $v$, the mean blueshift
observed for a line is $2v$/3.} is 6\%.  The intercepted flux is a
factor of two larger when we include the dipole nature of Thomson
scattering. Thus the total energy incident on the star is
$3\times10^{48}$ ergs.  This falls short of the required amount in GRB
011211, which is $50\zeta_{50}\times10^{48}E_{48}
=5\times10^{49}\zeta_{50}E_{48}$ ergs.  If the supernova explosion is
non-spherical, as evidenced by recent polarization observations, with
the equatorial region expanding at about twice the speed of the polar
region, then about 50\% of the back scattered flux will be intercepted
by the ejecta and the total energy incident on the expanding stellar
surface is $\sim 2\times 10^{49}$ ergs.  Instead of an asymmetric
supernova, we could also consider a rapidly expanding cocoon, possibly
associated with the GRB jet.  This will work equally well provided the
density is sufficiently high (eq 6) and the transverse velocity is
$\gta0.3c$.

The deceleration radius for the relativistic GRB ejecta, for the wind
density deduced above, is significantly smaller than the radius $r_1$
of the pair screen. Thus a substantial fraction of the energy of the
relativistically expanding GRB fireball will be converted to x-ray
radiation before the expanding material hits the pair screen, and this
energy will also be scattered by the pair screen and intercepted by
the supernova ejecta.  The early afterglow spectral peak is typically
at an energy smaller than the GRB peak, and the peak shifts to lower
energies with time as the Lorentz factor of the shock decreases. In
the case of the high density surrounding medium considered here, the
two emissions could have similar peak energy at early times, but the
afterglow emission will subsequently shift to lower frequencies.

When we include the scattered afterglow flux, we find that the
expanding supernova ejecta could receive a total of $\sim
5\times10^{49}$ ergs in 1--20 keV band.  The afterglow radiation will
be more efficient, compared to the GRB photons, in ionizing Si and
other atoms which have ionization energies of a few keV.  Such a
continuum x-ray flux can be efficiently processed into line radiation
as discussed in the Appendix, with about 2\% of the continuum energy,
or $\sim 10^{48}$ ergs, coming out in Si K$_\alpha$ photons.  We note
that, because of projection effects, an observer who is located in the
direction of the jet will infer a factor of 2 higher line luminosity
than the true isotropically averaged luminosity.  This gives a safety
factor of $\sim2$ in the above estimates.

Because the pair cloud is moving outward with a Lorentz factor
$\sim2$, its optical depth is likely to reduce substantially by the
time the line photons arrive at the cloud.  Therefore, the x-ray lines
will not be significantly attenuated by scattering off pairs.  Note in
addition that GRB radiation scattered off the pair cloud associated
with the counter-jet is likely to be blocked by the expanding
supernova ejecta (for the parameters we have considered, viz.,
$\theta_j=0.2$, transverse $v/c\gta0.3$).  Thus, that radiation will
not contribute to the observed continuum flux.

\section{Summary and Discussion}

The discovery of x-ray lines in the afterglow spectra of GRBs, hours
to days after the initial gamma-ray burst, poses severe problem for
theoretical models.  We have focused in this paper on the strong Si
line observed by Reeves et al. (2002) in GRB 011211, but our arguments
are valid more generally and apply, for instance, to the Fe line
observed in GRB 991216 by Piro et al. (2000). In the case of GRB
011211, we have the following constraints:

\noindent
1. The line-emitting gas must be moving towards the observer at an
average speed of $v\sim0.1c$ relative to the rest frame of the GRB.
This suggests that the gas is associated with ejecta from a supernova
explosion.

\noindent
2. The time delay of $\sim10^4$ s (measured in the source frame)
between the initial GRB and the later line emission constrains the
geometry of the source.  In models in which the lines are produced by
direct irradiation by GRB photons, the distance of the line-emitting
region from the central engine ($D$) and the lateral extent of the
irradiated region ($R$) are constrained by equation (4).  In indirect
irradiation models, in which the GRB radiation is first scattered off
a nearby screen, the distance to the cloud is constrained to be
$\sim10^{14.5}$ cm and $D$ is smaller $\sim10^{13.5}$ cm.

\noindent 
3. According to Reeves et al. (2002), GRB 011211 emitted $10^{48}$
ergs or more in each of five K$_\alpha$ lines.  This estimate of
the energy is conservative since there might well have been additional
line emission during the 11 hours between the time of the GRB and
when the x-ray observations began.  Regardless, the energy in the
lines is a substantial fraction of the total gamma-ray energy budget
of $\sim10^{51}$ ergs (for reasonable beaming angles).  Assuming the
lines were produced by photoionization, the source must have been very
efficient at converting continuum radiation into lines.

\noindent
4. In the x-ray observations, the 1--10 keV continuum flux was found
to be no more than a factor of 10 greater than the flux in each line.
This introduces two constraints on models.  First, the free-free
emission from the gas should not be too large relative to the line
emission.  This requires the composition of the gas to be several
times solar (Appendix A3, Reeves et al. 2002) or the temperature of
the gas to be less than a keV.  Second, as Lazzati et al. (2002) have
argued, the photoionizing radiation from the GRB which is the cause of
the line emission must not be visible to the observer.  The reason is
that the irradiating flux is estimated to be nearly 100 times larger
than the line flux even under optimum conditions, much in excess of
the observed continuum flux.  This constraint eliminates all models in
which the central engine puts out substantial energy for $10^4$ s or
longer.

\noindent
5. For optimum line emission, the ionization parameter $\xi$ has to
lie in a relatively narrow range: $\xi_3=\xi/10^3 \sim 0.1-$few.  This
condition introduces quite a strong constraint on the electron density
in the line-emitting gas, as shown by equation (3).

When combined together, these constraints make it very hard to find a
model that can explain the lines claimed by Reeves et al. (2002) in
GRB 011211.  We have considered in \S2 two varieties of models that
have been discussed previously in the literature.

In one type of model, one has a ``transmission geometry'' in which the
irradiation occurs on one surface of a slab of gas and the lines are
emitted from the other surface.  The shell model of Reeves et
al. (2002) is of this type.  Given the density constraint (3) and the
requirement that the Thomson optical depth through the slab not be
much greater than unity (in order to have any transmission at all), we
deduce an impossibly thin slab at quite a large radius from the GRB
engine.  We believe it is extremely hard to arrange such a thin shell.

In another type of model, one has a ``reflection geometry'' in which
both the irradiation and the line emission occur from the same surface
of the line-emitting region.  A good example of such a model is the
funnel model discussed in \S2.  In this case, the high density implied
by equation (3), coupled with the large length $D$ of the funnel,
imply an enormous amount of mass in the external medium.  The mass can
be reduced to something reasonable only by invoking highly contrived
conditions.

We have also proposed and discussed in some detail a third type of
model which appears to have a better chance of producing observable
x-ray lines in GRBs. In this model, gamma-rays from the GRB create a
pair screen temporarily at a distance of $10^{14}-10^{15}$ cm from the
central engine.  The screen scatters a substantial fraction of the
gamma-rays backward and thereby irradiates the expanding supernova
ejecta.  The lines are produced in the outer layers of the ejecta.  In
this model, the time delay between the GRB and the line emission
reflects the added path length that the irradiating photons travel
before they photoionize the line-emitting gas.  The line-emitting
region is significantly more compact than in the other two models.
Also, each line-emitting gas element experiences irradiation for
$10^4$ s and not just $10^2$ s as in the other models.  These
modifications enable this model to overcome the problems associated
with the other models.  Interestingly, even though irradiation occurs
for $10^4$ s, the model can satisfy the low continuum flux (constraint
4 above) because the irradiating photons move backwards, i.e., away
from the observer.  (The corresponding backward moving photons from
the counter-jet are likely to be hidden by the supernova ejecta, see
\S3.)

The pair screen scattering model satisfies most of the constraints
described above without requiring too much mass or too much energy or
an implausible geometry.  Moreover, the distance at which a pair
screen is likely to be formed around a GRB 
is just what is needed to explain the observed time delay.  In
addition, the model has the important virtue that the GRB and the
supernova occur simultaneously.  In our opinion, this is an
improvement over the shell and funnel models discussed above, both of
which require the supernova explosion to have occurred days to weeks
before the GRB explosion.

The main problem with the pair screen scattering model is that it has
difficulty explaining the total energy in the lines.  Because the
model involves an extra stage of scattering, the amount of irradiating
flux is reduced compared to the other models.  Therefore, in order to
produce as much as $10^{48}$ ergs per line, we need (i) very efficient
scattering in the pair screen, which implies order unity optical depth
and a reasonably small outward Lorentz $\gamma_\pm$ for the pairs, and
(ii) a large surface area for the supernova ejecta so that much of the
reflected gamma-rays may be intercepted and reprocessed into lines.
These requirements can be satisfied, as we have discussed in \S3, but
it requires pushing some parameters to their limit --- in particular,
we need a rather high mass loss rate from the pre-supernova star
within a few years of the explosion, and an asymmetric supernova with
a ratio of equatorial to polar velocity of about 2.

It is important to emphasize that, although we have concentrated on
GRB 011211, many of the constraints we have described in this paper
should be relevant for any detection of hydrogenic K$_\alpha$ lines of
Si, Fe and other elements, in GRB afterglows.  For a robust detection
of a line with current instruments, the line energy has to be of order
$10^{48}$ ergs and the continuum level has to be low, as perhaps in
GRB 011211.  Also, the detection is likely to be made hours to days
after the initial GRB.  Therefore, the problems we have emphasized
with explaining the lines in GRB 011211 will carry over to line
detections in other GRBs.

\bigskip\noindent
Acknowledgments: PK is indebted to Alin Panaitescu for valuable
discussions on the physics of pair screens surrounding gamma-ray
bursts, and to Gregory Shields and Craig Wheeler for numerous
discussions on radiative effects, supernovae and GRBs.  RN was
supported in part by NSF grant AST-9820686.

\bigskip
\medskip
\centerline{\bf Appendix A}
\smallskip
\centerline{\bf A1. X-ray Lines from a Reflecting Slab}
\medskip

The flux in K$_\alpha$ lines of an atomic species $A$ (Si XIV for
instance) from a photoionized slab of gas is given by
$$ 
f_\alpha = n_{A^+} n_e R_{rec} h\nu_\alpha \lambda, \eqno(\new)
$$ 
where $n_{A^+}$ is the number density of $A^+$ ions, $h\nu_\alpha$ is
the line photon energy,
$$
R_{rec}=5.2\times 10^{-14} z b^{1/2}\left[ 0.429 + 0.5 \ln b + 0.496/b^{1/3}
 \right] 
$$
is the recombination rate for hydrogenic atoms (Seaton, 1959),
$b=1.58\times 10^5 z^2/T ~{\rm s^{-1}}$, and $\lambda$ is the smaller
of the mean free paths for ionizing photons ($\lambda_{ion}$) and line
photons escaping from the region ($\lambda_{\alpha}$). The two mean
free paths are given by
$$ 
\lambda_{ion}^{-1} = \sum_{i=1}^{z_A} \sigma_{ion,i}(\nu_A) n_i,
$$
where $\sigma_{ion,i}(\nu_A)=7.9\times 10^{-18} z^{-2}
(\nu_A/\nu_i)^{-3}$cm$^2$ is the photoionization cross-section for
atomic species $i$ at the frequency $\nu_A$, the threshold ionization
frequency for atom $A$, and
$$ 
\lambda_\alpha^{-1} = \eta n_e \sigma_T + \sum_i \sigma_{ion,i}(\nu_\alpha) 
   n_i,
$$
where $\nu_\alpha$ is the frequency of the line photon, $\eta\approx (\delta\nu/
\nu_\alpha)(c/v_e)$ is the maximum allowed optical depth to Thomson scatterings
such that the line broadening does not exceed the observed linewidth
$\delta\nu$, and $v_e$ is the electron thermal velocity.
The ionization fraction of the atom $A$ (assumed to be in the hydrogenic state)
is determined from the following ionization equilibrium equation
$$ 
n_{A^+} n_e R_{rec} = n_A \int d\nu\, {\sigma_{ion,A}(\nu)f_\nu\over h\nu}. 
$$

We have solved the above equations numerically.  The results are shown
in Fig. 1.  In the calculations, the effect of resonant line trapping
was included (see the discussion in A2 below), but the Auger effect
was not considered.  The temperature was taken to be 10$^6$ K for all
the calculations, and the spectrum of the illuminating x-ray continuum
was taken to be $\nu^{-1.25}$.  For $T=10^7$ K, the peak of the Si
line flux shifts to $\xi\sim400$ for solar abundance and to
$\xi=2\times 10^3$ for 10 times solar abundance, and the value of
$f_{Si}/f_{inc}$ at the peak remains nearly the same. For the Fe XXIV
line, the line emission efficiency begins to decline at $2\times10^3$
for $T=10^7$K. For $T=4\times 10^7$K our results agree with those of
Lazzati et al. (2002) despite several differences in the input
physics.

We discuss here an approximate analytical estimate for the line flux
which provides some insight into the numerical results.  Since the
ionization cross-section decreases rapidly with increasing frequency,
as $\nu^{-3}$, the above equation can be written approximately as
$$ 
n_{A^+} n_e R_{rec} \sim n_A \sigma_{ion,A}(\nu_A) f_{\nu_A}/3h
   \sim {f_{\nu_A}\over 3h\lambda_{ion,A}},
$$
where $\lambda_{ion,A}^{-1}=n_A \sigma_{ion,A}(\nu_A)$ is the
photo-absorption depth for photons of frequency $\nu_A$ (the threshold
frequency for ionizing $A$). Substituting this in equation (7) we find
$$ 
f_\alpha \approx \nu_A f_{\nu_A} \left({\nu_\alpha\over 3\nu_A}\right)
 \left({\lambda\over\lambda_{ion,A} } \right).
$$

The photoionization cross-section for Si XIV is
$4\times10^{-20}$ cm$^2$ or a factor of $6\times10^4$ larger than the
Thomson cross-section. The abundance of Si, by number, for a gas of
solar composition is a factor of about $3\times10^4$ smaller than the
hydrogen abundance. Thus, $\lambda$ is approximately equal to the
Thomson mean free path ($\lambda_T$) for solar abundance and for an
ionization fraction of $\sim 50$\%.

For small values of the ionization parameter ($\xi\ \lta\ 10^2$) the ionization
fraction is small and the mean free path $\lambda\approx \min\{\lambda_{ion,A},
\lambda_\alpha\}\approx\lambda_\alpha$. In this case $f_\alpha/\xi$ 
increases with increasing $\xi$ because line photons are less likely
to be absorbed, as the neutral fraction of atoms of $z<z_A$ decreases
with increasing $\xi$.  Resonant line scattering makes this effect
very important.  At some value of $\xi\sim 10^2$, when
$n_{SiXV}/n_{SiXIV}\sim 1/2$ (ionization fraction of 50\%) the various
mean free-paths are equal to each other and the line is most
efficiently generated, i.e., $f_\alpha/\xi$ attains its maximum
value. At larger $\xi$ atoms are photoionized to larger depth but the
line photons can only come from a smaller depth of order $\lambda_T$.
Thus, $f_\alpha/\xi$ decreases roughly as $\xi^{-1}$.

The largest possible value for the line flux is $\sim\nu_\alpha
f_{\nu_A}/3$.  This flux is obtained under the right conditions when
the abundance of $A$ is larger by an order of magnitude than the
abundances of lower atomic number elements ($z_A-5\ \lta\ z<z_a$). It
is also clear that increasing the abundance of all metals, or even all
neighboring $z$ elements within $|z-z_A|\ \lta\ 5$, will not increase
the line emission efficiency $f_\alpha/\xi$ since, near the maximum of
$f_\alpha/\xi$, $\lambda$ is largely controlled by the photoionization
depth, which decreases inversely with increasing abundance.

Under optimal conditions, about 5\% of the 1-10 keV continuum flux can
be converted to line emission. This condition is met when the peak of
the incident radiation is close to $\nu_A$ and the atomic abundance of
$A$ is larger by a factor of a few than that of neighboring z-elements

\bigskip
\centerline{\bf A2. Resonant Line Trapping}
\medskip

The scattering cross-section for K$_\alpha$ photons of a hydrogenic atom
is
$$ 
\sigma_{21}(\nu) = {c^2\over 2\pi \nu_\alpha^2} {1\over
    1 + [4\pi(\nu-\nu_\alpha)/A_{21}]^2},
$$
where $A_{21}\approx 10^9 z^4$ s$^{-1}$ is the spontaneous decay
rate from $n=2$ to $n=1$. The mean scattering cross-section within
the Doppler width of the line is $\sim A_{21} c^2/(8\pi^2 \nu_\alpha^2
\delta\nu_D)$. The Doppler shift is $\delta\nu_D \approx
(3kT/2m_p z)^{1/2}\nu_\alpha/c=3\times 10^{-4} T_7^{1/2} \nu_\alpha
(15/z)^{1/2}$, and therefore the scattering cross-section, within the
Doppler width of the line, is $\sim 3\times 10^{-18}$ cm$^2$, or two
orders of magnitude larger than the photoionization
cross-section. Since the photoionization optical depth of the region
from which line photons are emerging is order unity, we see that the
scattering optical depth is $\sim10^2$.  Thus the probability that the
K$_\alpha$ photons are absorbed by atoms of lower atomic number, by
photoionizing them, on their way out of the line producing region is
not small and must be included in the calculation of the line flux.

It is interesting to note that a K$_\alpha$ photon undergoes only
$\sim \tau_\alpha^{2/3}$ scatterings before getting out of the
trapping region (and not $\tau_\alpha^2$ scattering, as would be the
case if the scattering cross-section was frequency independent), and
the total path length traversed before escape is $\sim
\tau_\alpha^{1/3}$ times the depth $d$ of the medium.  This is
explained below.

In each scattering the frequency of a line photon undergoes a random walk.
After $N$ scatterings the frequency shift is given by
$$
(\delta\nu/\nu_\alpha)^2 = N(v/c)^2. 
$$
Since the scattering cross-section for a Lorentzian profile 
decreases as 
$(A_{21}/\delta\nu)^2 \propto (c/v)^2/N$, after $\tau_\alpha$
scatterings the cross-section decreases by a factor of $\tau_\alpha$.
The photon then finds itself in a region with optical depth of order
unity and escapes.  Taking into account the distance traveled toward
the surface in $N$ scatterings, which is proportional to $N^{3/2}$, we
find that photons escape after $\tau_\alpha^{2/3}$ scatterings and
therefore the total path length traversed by a photon is about
$\tau_\alpha^{1/3}d$ (where $d$ is the depth of the line producing
region).

When the absorption mean free path of the photon is comparable to the
photoionization depth for Si atoms, the line-photon flux will be
suppressed by a factor of about $\tau_\alpha^{1/3}$. For an ionization
parameter of $400$ and $T_7=0.1$, or for $\xi=100$ and $T_7=1$, the
two mean free paths are equal and the line flux is small. However, as
we increase $\xi$ the neutral atomic fraction decreases and the path
length over which a line photon is absorbed by neutral atoms of lower
atomic number goes up. An increase in the ionization parameter by a
factor of four increases the photo-absorption length by a factor of
four, and the emergent line emission is no longer suppressed by line
trapping.

The enhanced photo-absorption of line photons due to resonant line
trapping is included in the numerical results shown in Fig. 1.  Our
basic result is that the maximum flux in the Si XIV K$_\alpha$ line is
a few percent of the irradiating x-ray continuum flux, and that the
maximum flux is attained when $\xi\sim 10^3$ for solar composition and
for $T=10^6$K (for $T=4\times10^7$K, $\xi\sim100$ as reported in
Lazzati et al., 2002). The optimum $\xi$ increases further if the Si
abundance is above solar.

\medskip
\centerline{\bf A3. Bremsstrahlung Energy Loss}
\medskip

Let us assume that the most abundant ionic species in the plasma has
an atomic number $z$. Its number density compared to electrons is
smaller by a factor of $z$, and the bremsstrahlung energy loss rate
per electron in the gas is given by
$$ 
\epsilon_B = 5.3\times10^{-24} n_e z T_7^{1/2} \;\;{\rm ergs\, s^{-1}}.
$$
If we approximate the geometry of the region as locally
plane-parallel, then the x-ray continuum flux due to bremsstrahlung is
given by
$$
f_B \approx \epsilon_B n_e \lambda_T = \epsilon_B/\sigma_T = 8.0\times10^{15} 
n_{15} T_7^{1/2} z ~{\rm ergs\,cm^{-2}s^{-1}}.
$$
Comparing this with the estimate for the Si line flux given in
equation (2), we find (for $z=z_{14}=\tau=\epsilon_3=1$)
$$
{f_B\over F_{Si}}\sim 26a_s.
$$
Observations of GRB 011211 indicate that the continuum flux is no more
than ten times the Si line flux.  This suggests that the line-emitting
gas is overabundant in heavy elements by a factor of at least a few.
Reeves et al. (2002) estimated an abundance of ten times solar in
their analysis.  Alternatively, the temperature of the gas must be
less than 1 keV, so that the bulk of the bremsstrahlung emission is
outside the 1--10 keV band.

The ionization parameter $\xi_B$ corresponding to the above
bremsstrahlung flux is
$$ 
\xi_B = {4\pi f_B\over n_e} = 100 { T_7^{1/2} z}.\eqno(\new)
$$
For a pure H plasma this gives an ionization parameter of 100. If the
ionization parameter for the incident radiation is $50\zeta_{50}$,
then the bremsstrahlung flux will equal the incident x-ray flux, which
is ruled out by the above-mentioned limit on the continuum flux.  We
need to lower the bremsstrahlung flux by a factor of about 10 to
satisfy the continuum limit.  From equation (\last) this requires
reducing the temperature to $T_7\sim 0.1$.  Lowering the temperature
increases the recombination rate, and in order to get to the same
ionization fraction we must increase the ionization parameter by a
factor of $T_7^{-1/2}$.  The upshot is that in order to get about
2-4\% of the irradiating flux as Si line emission, without violating
the constraint on the x-ray continuum radiation, we require the
ionization parameter to be $\sim 10^3$. The argument presented in A2
also suggests $\xi\sim 10^3$ in order that K$_\alpha$ photons are not
absorbed by atoms of lower $z$.

\vfill\eject

\vfill\eject

\begin{figure}
\plotone{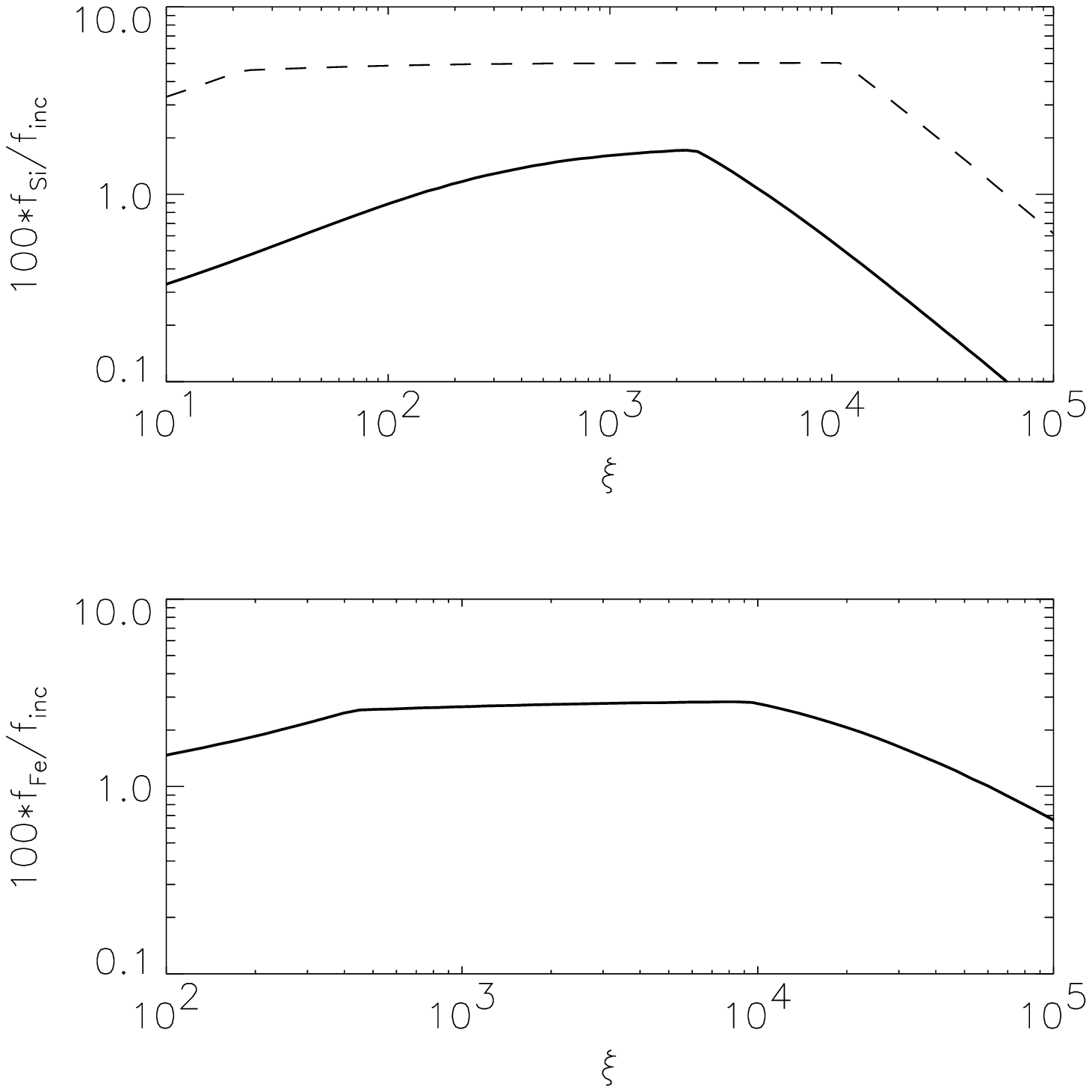}
\caption{ The upper panel shows the flux in the Si XIV K$_\alpha$ line
divided by the incident flux in the 1--15 keV band as a function of
the ionization parameter $\xi$.  The solid line corresponds to solar
composition and the dashed line to when the abundance of Si alone is
taken to be ten times the solar value but the abundances of other
elements are kept at solar.  If the abundances of all metals are
increased by the same factor, the curve would be similar to the solid
line for $\xi\ \lta\ 2000$, but the decline of $f_{Si}/f_{inc}$ at
large $\xi$ would start at a larger value of $\xi$ (as in the
corresponding results shown in Lazzati et al. 2002). The lower panel
shows the flux in the Fe XXIV K$_\alpha$ line divided by the x-ray
continuum flux.}
\end{figure}

\end{document}